\documentclass[twocolumn,tighten,twocolappendix]{aastex6}

\usepackage{amsmath}
\usepackage{graphicx}
\usepackage{epstopdf}
\usepackage{amssymb}
\usepackage{extarrows}
\usepackage{color}
\usepackage{bm}

\def\be{\begin{equation}}
\def\ee{\end{equation}}
\def\ba{\begin{eqnarray}}
\def\ea{\end{eqnarray}}
\def\go{\mathrel{\raise.3ex\hbox{$>$}\mkern-14mu
             \lower0.6ex\hbox{$\sim$}}}
\def\lo{\mathrel{\raise.3ex\hbox{$<$}\mkern-14mu
             \lower0.6ex\hbox{$\sim$}}}

\newcommand{\bl}{\bm {\hat l}}
\newcommand{\Dl}{\Delta{\bm l}}
\newcommand{\bL}{\bm L}
\newcommand{\der}{\text{d}}

\newcommand{\omjd}{\omega_{jd}}
\newcommand{\Omdp}{\Omega_{dp}}
\newcommand{\Omjp}{\Omega_{jp}}
\newcommand{\hOmdp}{\hat\Omega_{dp}}
\newcommand{\hOmjp}{\hat\Omega_{jp}}
\newcommand{\thjd}{\theta_{jd}}

\newcommand{\phjd}{\varphi_{jd}}

\newcommand{\bY}{\bm Y}
\newcommand{\bA}{\bm A}
\newcommand{\bB}{\bm B}

\def\red#1 {\textcolor{red}{#1}\ }   
\def\blue#1 {\textcolor{blue}{#1}\ }   

\begin{document}

\title{Hiding Planets Behind a Big Friend: Mutual Inclinations of
Multi-Planet Systems with External Companions}


\author{Dong Lai and Bonan Pu}
\affil{Cornell Center for Astrophysics and Planetary Science, Department of Astronomy,
Cornell University, Ithaca, NY 14853, USA}

\begin{abstract}
The {\it Kepler} mission has detected thousands of planetary systems
with 1-7 transiting planets packed within 0.7~au from their host
stars. There is an apparent excess of single-transit planet systems
that cannot be explained by transit geometries alone, when a single
planetary mutual inclination dispersion is assumed. This suggests that
the observed compact planetary systems have at least two different
architectures.  We present a scenario where the ``Kepler dichotomy''
may be explained by the action of an external giant planet or
stellar companion misaligned with the inner multi-planet system.  The
external companion excites mutual inclinations of the inner
planets, causing such systems to appear as ``Kepler singles'' in
transit surveys.  We derive approximate analytic expressions (in
various limiting regimes), calibrated with numerical calculations, for
the mutual inclination excitations for various planetary systems and
perturber properties (mass $m_p$, semi-major axis $a_p$ and inclination
$\theta_p$). In general, the excited mutual inclination increases with
$m_p/a_p^3$ and $\theta_p$, although secular resonances may lead to
large mutual inclinations even for small $\theta_p$.
We discuss the implications of our results for understanding the
dynamical history of transiting planet systems with known external perturbers.
\end{abstract}

\keywords{planetary systems --- planets and satellites: dynamical
  evolution and stability --- planets and satellites: formation
  --- stars: individual (Kepler-48, Kepler-56, Kepler-68, Kepler-444, 
  Kepler-454, WASP-47, GJ 832, 55 Cancri)}

\section{Introduction}

NASA's {\it Kepler} mission has discovered $\sim 4700$ planet
candidates (as of May 2016), about half of which are confirmed planets
(e.g. Mullally et al.~2015; Coughlin et al.~2016; Morton et al.~2016).
Most of these planets are super-Earths or sub-Neptunes (with radii
1.2-3$R_\oplus$), and have orbital periods less than 200~days.  Among
the $n_{\rm tran}=3606$ Kepler planetary systems, $80\%$ have one
transiting planet, and $20\%$ have 2-7 transiting planets
[The number of systems with $N_{\rm tran}$ planets is $n(N_{\rm
    tran})=2871,\,492,\,158,\,61,\,20,\,3,\,1$ for $N_{\rm
    tran}=1,2,\cdots,7$].\footnote{Data was retrieved form the NASA
  Exoplanet Archive on May 17, 2016; planets with a KOI deposition
  ``False Positive'' were removed from this sample.}  The observed
transit multiplicity distribution, $f(N_{\rm tran})=n(N_{\rm
  tran})/n_{\rm tran}$, and its dependence on the sizes and periods of
planets, contain useful information on the architecture of compact
planetary systems, such as the true planet multiplicity,
the mutual inclinations, and orbital spacings between adjacent
planets.  In general, there exists a degeneracy between these
(underlying) quantities in producing the same $f(N_{\rm tran})$.  For
example, larger planet spacings and mutual inclinations will raise the
relative number of single-transit systems (Lissauer et al.~2011;
Tremaine \& Dong 2012). This degeneracy can be partially lifted by
combining the statistics of $f(N_{\rm tran})$ with the result of RV
surveys (Tremaine \& Dong 2012; Figueira et al.~2012), or using the
transit duration ratios of different planets orbiting the same star
(Fabrycky et al.~2014). The general conclusion from a number of
studies is that Kepler compact planetary systems are flat, with the
inclination dispersion of order a few degrees (Lissauer et al.~2011;
Tremaine \& Dong 2012; Figueira et al.~2012; Johansen et al.~2012;
Fang \& Margot 2012; Fabrycky et al.~2014).

It has been noted that models with a single mutual inclination
dispersion (e.g. in a Rayleigh distribution) fall short in explaining
the large number of single-transit ($N_{\rm tran}=1$) systems relative
to multiple-transit (higher-$N_{\rm tran}$) systems by a factor of two
or more (Lissauer et al.~2011; Johansen et al.~2012; Weissbein et
al.~2012; Ballard \& Johnson 2016).
\footnote{This result depends somewhat
on the assumed forms of the underlying multiplicity function 
(Tremaine \& Dong 2012), since there is a degeneracy between the mutual inclination
distribution and the multiplicity function.} 
This suggests that the Kepler
planetary systems may consist of at least two underlying populations with
different architectures: The first has many ($\go 6$) planets with
small ($\lo 2^\circ$) mutual inclinations, and accounts for the
majority of the $N_{\rm tran}\ge 2$ systems; the second has fewer
planets or higher mutual inclinations, and accounts for a significant
portion of the observed single-transit systems. This is the so-called
``Kepler Dichotomy''. Xie et al.~(2014) found that the multi-transit
systems are more likely to exhibit detectable transit timing
variations than the single-transit systems, suggesting that the former
are more closely packed than the latter.
Morton \& Winn (2014) found that the obliquities of stars with a
single transiting planet are systematically larger than those with
multiple transiting planets (see Albrecht et al.~2013), again suggesting that a substantial
fraction of Kepler's single-transit systems are dynamically hotter
than the flat multiple-transit systems.

The origin of the Kepler dichotomy is unknown. The observed Kepler
multi-planet systems appear to be tightly packed and close to the edge
of instability (Fang \& Margot 2013; Pu \& Wu 2015; Volk \& Gladman
2015).  Thus a dichotomy in planetary architectures may arise from the
long-term evolution of dynamically full systems. In this picture, the
more densely packed systems underwent dynamical instability, leading
to planet collision/consolidation and the formation of Kepler
``singles'' (Pu \& Wu 2015; Volk \& Gladman 2015). It is unclear
to what extent dynamical instability can account for the Kepler
dichotomy quantitatively, as the observed Kepler multi's are
sufficiently ``cold'' and not massive enough to experience appreciable
inclination excitation or dynamical instability within the stellar
lifetime (Johansen et al.~2012; Becker \& Adams 2016).
On the other hand, the Kepler dichotomy may have a primordial origin,
and results from the in-situ assembly of planetesimal disks (Hansen \& Murray 2013) 
with different masses and density profiles (Moriarty \& Ballard 2015).

In this paper we study the excitation of mutual inclinations in a
compact multi-planet system by an external giant planet or stellar companion
(Sections 2 and 3). 
In general, the giant planet may be on a misaligned orbit relative to the
inner planetary system, as a result of warp in protoplanetary disks
(e.g., Foucart \& Lai 2011,2014) or strong scatterings between
multiple giants (Juric \& Tremaine 2008; Chatterjee et al.~2008).
{A distant stellar companion may also be inclined because of its 
misaligned orbital angular momentum at birth (e.g. Hale 1994).}
By exciting mutual inclinations of the inner planets, the giant planet
can ``heat up'' the inner multi-planet system, causing it to appear as
a single-transit system.

{Since $\sim 50\%$ of the solar-type stars are in binaries, it is not
surprising that many exoplanetary systems (including Kepler planet
candidates) have been found to have external binary companions with a
range of separations (e.g., Baranec et al.~2016). There is
observational evidence that relatively close-by stellar companions
(with separation $\lo 20-50$~au) tend to reduce the planet formation
efficiency (e.g., Wang et al.~2015a; Kraus et al.~2016; Ngo et
al.~2016).  Wang et al.~(2015b) found that $5\pm 5\%$ of Kepler
multi's have stellar companions at separation 1-100~au, compared to
$21\%$ for field stars in the solar neighborhood, suggesting that such
companions can misalign or disrupt multi-planet systems.}
On the other hand,
RV surveys continue to reveal a population of giant planets at large distances
($\go$ a few au) from their host stars (e.g., Marmier et al.~2013; Feng et al.~2015;
Moutou et al.~2015; Rowan et al.~2016; Wittenmyer et al.~2016;
Bryan et al.~2016). 
The Keck survey suggests that about $20\%$ of solar-type stars
could host gas giants within 20 au (Cumming et al.~2008), while HARPS finds that 
$14\%$ of such stars host giant planets with periods less than 10 years.
Because of the limited time span and the faint magnitudes of Kepler stars,
the current census of distant giant companions to Kepler compact systems is rather 
incomplete. Nevertheless, a number of such long-period companions or candidates
have been found using the transit method (Schmitt et al.~2014; 
Uehara et al.~2016) and the RV method (e.g., Kepler-48, Kepler-56, Kepler-68,
Kepler-90, Kepler-454);
a number of non-{\it Kepler}
``inner compact planets + giant companion'' systems have also been found
(e.g., GJ~832, WASP-47)
-- see Section 4 for applications of our theory to some of these systems.
Bryan et al.~(2016) reported that about $50\%$ of one and two-planet systems
discovered by RV have companions in the 1-20$M_J$ and 5-20~au range.
All these results indicate that external ($\go 1$~au) giant planet companions 
are common around hot/warm ($\lo 1$~au) planets, and may significantly shape 
the architecture of the inner planetary systems.

We note that the possible role of external companions on compact
planetary systems has often been noted (e.g. Lissauer et al.~2011) and
formal secular theories (with various approximations) suitable for
such study have been presented before (e.g. Tremaine et al.~2009; Boue \& Fabrycky 2014).
Our paper makes progress on this problem by deriving simple
approximate analytic expressions (in various limiting regimes),
calibrated with numerical results (Sections 2 and 3), that allow us to answer the
question: Given an inner planetary system, what are the mutual
inclinations excited by an external perturber of mass $m_p$, semi-major
axis $a_p$ and inclination $\theta_p$? In general, a ``strong'' perturber
(with large $m_p/a_p^3$) with high $\theta_p$ leads to 
larger mutual inclinations in the inner planets. But our work also 
reveals that under some conditions, large mutual inclinations can be 
generated even for small $\theta_p$ ($\lo 1^\circ$) because of
secular resonances.



\section{Two-Planet Systems with External Perturber}

Consider two planets (mass $m_1$ and $m_1$) in circular orbits
(semi-major axes $a_1$ and $a_2$, with $a_2>a_1$) around a central star
(mass $M_\star$). The two planets are initially coplanar.  An external
perturber (mass $m_p$) moves in a circular inclined orbit
\footnote{When the perturber has an finite eccentricity $e_p$,
we can simply replace $a_p$ by $a_p\sqrt{1-e_p^2}$ in all equations
to capture the leading quadrupole-order effect of the perturber on the planets
(e.g., Liu et al.~2015).},
with semi-major axis $a_p$ ($>a_1,a_2$) and inclination angle $\theta_p$.
How does the mutual inclination of the two inner planets evolve?

We denote the angular momentum vectors to the three planets by 
$\bL_1=L_1\bl_1$, $\bL_2=L_2\bl_2$ and $\bL_p=L_p\bl_p$, where 
$\bl_1$, $\bl_2$ and $\bl_p$ are unit vectors. When $L_p\gg L_1,~L_2$, the unit vector 
$\bl_p$ is fixed in time. The evolution of $\bl_1$, $\bl_2$ is governed by
\ba
&&{\der\bl_1\over \der t}=\omega_{12}(\bl_1\cdot\bl_2)(\bl_1\times\bl_2)
+\Omega_{1p}(\bl_1\cdot\bl_p)(\bl_1\times\bl_p),\label{eq:dl1}\\
&&{\der\bl_2\over \der t}=\omega_{21}(\bl_1\cdot\bl_2)(\bl_2\times\bl_1)
+\Omega_{2p}(\bl_2\cdot\bl_p)(\bl_2\times\bl_p).\label{eq:dl2}
\ea
Here $\omega_{12}$ measures the precession rate of $\bl_1$ around
$\bl_2$ (driven by $m_2$), and $\Omega_{1p}$ the precession rate of $\bl_1$ around $\bl_p$
(driven by $m_p$):
\ba
&&\omega_{12}={Gm_1m_2a_1\over 4a_2^2L_1} b_{3/2}^{(1)}\!\left({a_1\over a_2}\right),
\label{eq:o12}\\
&&\Omega_{1p}={Gm_1m_pa_1\over 4a_p^2L_1} b_{3/2}^{(1)}\!\left({a_1\over a_p}\right),
\label{eq:O1p}
\ea
where $b_{3/2}^{(1)}(\chi)$ is the Laplace coefficient:
\be
b_{3/2}^{(1)}(\chi) = \frac{2}{\pi} \int_0^\pi \frac{\cos\phi\, \der\phi}
{(1- 2\chi \cos\phi + \chi^2)^{3/2}}.
\ee
Similar expressions apply to $\omega_{21}$ and $\Omega_{2p}$. Clearly
\ba
&&{\omega_{21}\over\omega_{12}}={L_1\over L_2}={m_1\over m_2}\left({a_1\over a_2}\right)^{1/2},\\
&&{\Omega_{2p}\over\Omega_{1p}}=\left(\!{a_2\over a_1}\!\right)^{\!1/2}
{b_{3/2}^{(1)}(a_2/a_p)\over b_{3/2}^{(1)}(a_1/a_p)}.
\label{eq:Omratio}\ea
Note that Eqs.~(\ref{eq:dl1})-(\ref{eq:O1p}) are approximate but
become exact in two limiting cases: (i) $\bl_1$, $\bl_2$ and $\bl_p$
are nearly aligned (e.g., Tremaine 1991); (ii) $\chi\ll 1$, in which
case the quadrupole approximation is accurate and $b_{3/2}^{(1)}(\chi)
=3\chi [1+(15/8)\chi^2+(175/64)\chi^4+\cdots]\simeq 3\chi$ (Murray \& Dermott 1999).

\subsection{Numerical Result}

\begin{figure}
\centering
\includegraphics[scale=0.28]{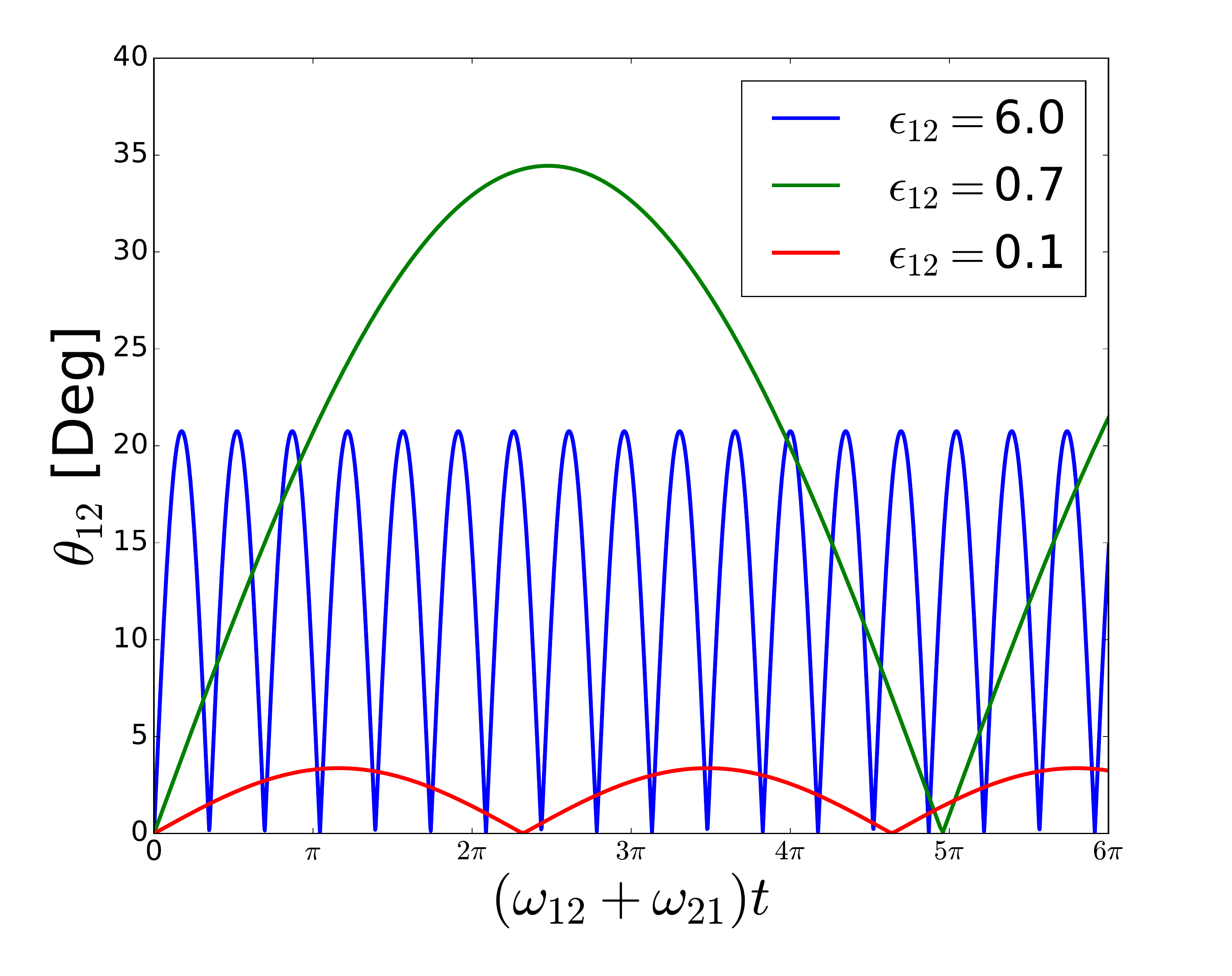}
\caption{Time evolution of the mutual inclination angle between 
two inner planets in the presence of an external inclined perturber
(at $\theta_p=10^\circ$). The two planets have mass ratio
$m_2/m_1=10$, with $a_1=0.3$~au, $a_2=0.5$~au and are initially aligned.
Different curves are for different values of $\epsilon_{12}$, corresponding
to different strengths of the perturber ($m_p/a_p^3$). For $\epsilon_{12}\ll 1$,
$\theta_{12}$ oscillates with the characteristic frequency $(\omega_{12}+\omega_{21})$
(see Eq.~\ref{eq:os1}); for $\epsilon_{12}\gg 1$, the characteristic frequency is
$(\Omega_{2p}-\Omega_{1p})\cos\theta_p$ (see Eq.~\ref{eq:osc2});
near the resonance ($\epsilon_{12}\sim 1$; see the green curve), 
the characteristic frequency is much smaller.}
\label{fig1}
\end{figure}

\begin{figure}
\centering
\includegraphics[scale=0.31]{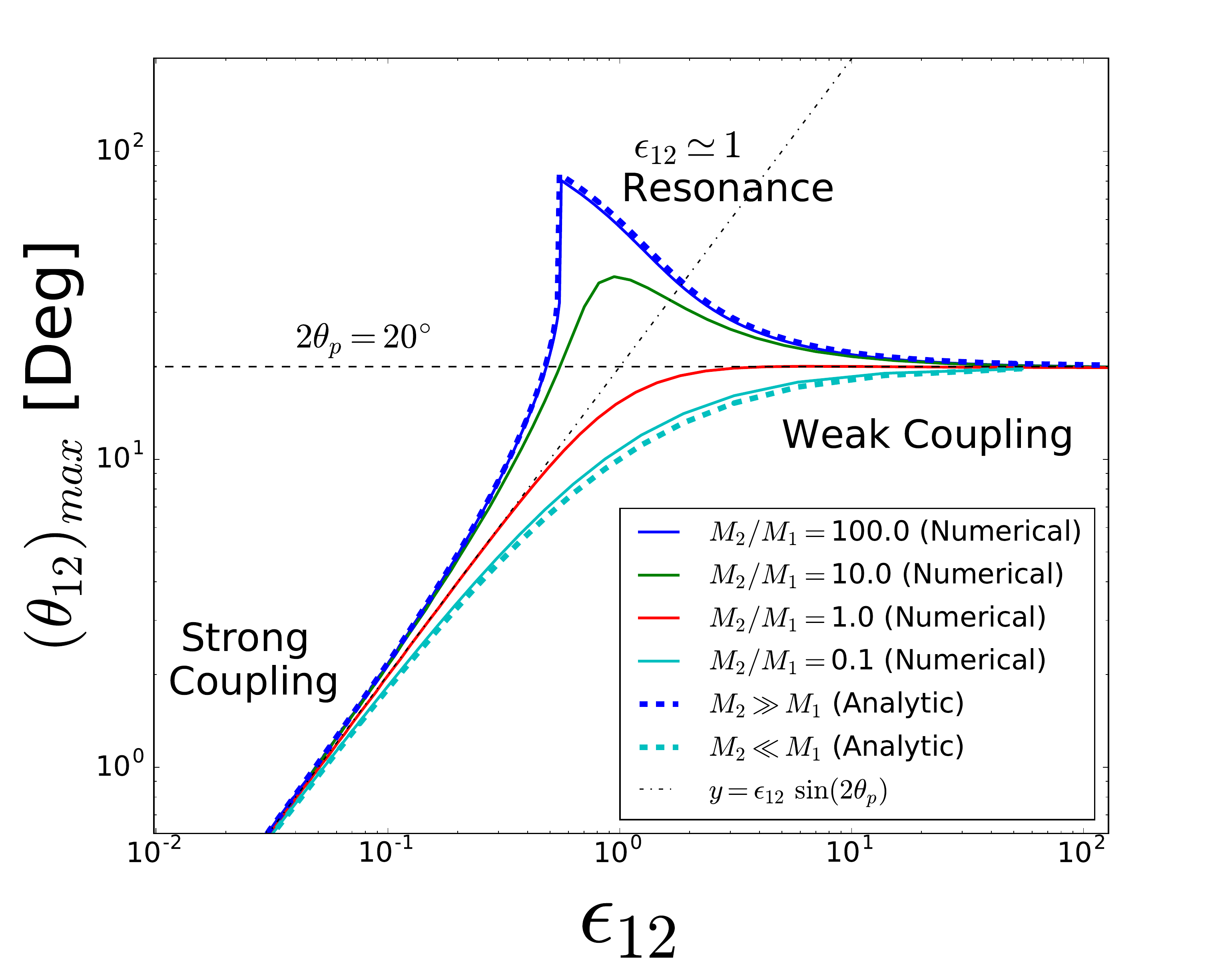}
\caption{Maximum mutual inclination between two inner planets
($m_1,m_2$) in the presence of an external perturber ($m_p$).
The inner planets are initially coplanar ($\theta_{12}=0$)
and inclined relative to the perturber at $\theta_p=10^\circ$.
The two planets are located at $a_1=0.3$~au and $a_2=0.5$~au 
(but only $a_2/a_1$ affects the result).
The different curves correspond to different mass ratios, as indicated. 
The dimensionless parameter $\epsilon_{12}$ 
(see Eq.~\ref{eq:eps12} or \ref{eq:epsilon12}) is varied by varying the
``strength''of the perturber, $m_p/a_p^3$. Analytical results in the strong coupling 
and weak coupling limits are also shown. A resonance feature is present when $m_2/m_1\go 1$.
}
\label{fig2}
\end{figure}

We integrate Eqs.~(\ref{eq:dl1})-(\ref{eq:dl2}) with an initially aligned pair of 
inner planets, and an inclined external perturber at $\theta_p=10^\circ$.
Figure 1 shows a few examples of the time evolution of the mutual inclination
angle ($\theta_{12}$) between the two inner planets, for 
$a_1=0.3$~au, $a_2=0.5$~au, $m_2/m_1=10$, and several values of $\epsilon_{12}$, 
as defined by 
\be 
\epsilon_{12}\equiv
{\Omega_{2p}-\Omega_{1p}\over\omega_{12}+\omega_{21}}.  
\label{eq:eps12}
\ee
Figure 2 depicts the maximum mutual inclination, $(\theta_{12})_{\rm max}$, 
as a function of $\epsilon_{12}$ for several different values of 
mass ratio $m_2/m_1$.
The dimensionless parameter $\epsilon_{12}$ measures whether
the inner planets are strongly coupled ($\epsilon_{12}\ll 1$) or
weakly coupled ($\epsilon_{12}\gg 1$); see below. Note that
$\epsilon_{12}$ can be written as
\be
\epsilon_{12}={\hat\Omega}_{1p}\,\, {(\Omega_{2p}/\Omega_{1p})-1\over
1+(L_1/L_2)}.
\ee
Here $L_1/L_2=(m_1/m_2)(a_1/a_2)^{1/2}$ is the ratio of the planet's angular momenta ,  
$\Omega_{2p}/\Omega_{1p}$
is given by Eq.~(\ref{eq:Omratio}), which simplifies to 
\be
{\Omega_{2p}\over\Omega_{1p}}
\approx \left({a_2\over a_1}\right)^{3/2}
\ee
in the quadrupole approximation (valid for $a_p\gg a_2,a_1$), and 
\ba
{\hat\Omega}_{1p} &\equiv & {\Omega_{1p}\over\omega_{12}}=
{m_p\over m_2}\,{a_2^2\, b_{3/2}^{(1)}(a_1/a_p)\over  a_p^2\, b_{3/2}^{(1)}(a_1/a_2)}
\nonumber\\
&\approx & {m_p\over m_2}\left({a_2\over a_p}\right)^3
\,{3 a_1/a_2\over  b_{3/2}^{(1)}(a_1/a_2)},
\ea
where the last equality assumes $a_p\gg a_1$.
Thus
\be
\epsilon_{12}\approx \left({m_p\over 10^3m_2}\right)\left({10a_2\over a_p}\right)^3
\left[{3 a_1/a_2\over  b_{3/2}^{(1)}(a_1/a_2)}\right]\,{(a_2/a_1)^{3/2}-1\over 1+(L_1/L_2)}.
\label{eq:epsilon12}\ee
For given inner planet parameters ($m_{1,2}$ and $a_{1,2}$ -- in fact, only the ratios
$m_2/m_1$ and $a_2/a_1$ matter),
the result for $(\theta_{12})_{\rm max}$ depends on $\theta_p$ and on 
$m_p$ and $a_p$ through the combination $m_p/a_p^3$ (for $a_p\gg a_1,a_2$).

In the following subsections we discuss the the behaviors of
$(\theta_{12})_{\rm max}$ in the limits of $\epsilon_{12}\ll 1$
(strong coupling) and $\epsilon_{12}\gg 1$ (weak coupling), and as
well as the resonance feature around $\epsilon_{12}\sim 1$.

\subsection{Strong Coupling Limit: $\epsilon_{12}\ll 1$}

In the {\it strong coupling limit},
with $\epsilon_{12}\ll 1$ (see Eq.~\ref{eq:eps12}), we expect $\bl_1$ and $\bl_2$
to stay close to alignment. Let $\bL=\bL_1+\bL_2\equiv L\bl$ be the
total angular momentum of the two inner planets, with $L\simeq L_1+L_2$. 
From Eqs.~(\ref{eq:dl1})-(\ref{eq:dl2}), we find
\be
{\der\bl\over \der t}\simeq \Omega_L(\bl\cdot\bl_p)(\bl\times\bl_p),
\label{eq:dL}
\ee
where $-\Omega_L(\bl\cdot\bl_p)$ is the precession rate of $\bl$ around $\bl_p$, with
\be
\Omega_L\simeq {L_1\Omega_{1p}+L_2\Omega_{2p}\over L_1+L_2}.
\ee
In the frame corotating with $\bl$, we have
\ba
&&\biggl({\der\bl_1\over \der t}\biggr)_{\rm rot}=
{\der\bl_1\over \der t}+\Omega_L (\bl\cdot\bl_p)(\bl_p\times \bl_1)\nonumber\\
&&\simeq \left[\Bigl(\Omega_L \bl\!\cdot\bl_p-\Omega_{1p} \bl_1\!\cdot\bl_p
\Bigr)\bl_p-\omega_{12}(\bl_1\!\cdot\bl_2) \bl_2\right]\times\bl_1.
\label{eq:dlrot}
\ea
Let $\bl_{1,2}=\bl+\Dl_{1,2}$, with $|\Dl_{1,2}|\sim \epsilon_{12} \ll 1$. Note that 
\be
L_1\Dl_1+L_2 \Dl_2\simeq 0.
\label{eq:l1l2}
\ee
Equation (\ref{eq:dlrot}) then becomes, to leading order in $\epsilon_{12}$,
\be
\biggl({\der\Dl_1\over \der t}\biggr)_{\rm rot}\simeq -(\omega_{12}+\omega_{21})\bl\times\Dl_1
+(\Omega_{1p}-\Omega_L)(\bl\cdot\bl_p)(\bl\times\bl_p),
\ee
For $\Dl_1(t=0)=0$, the leading-order solution is 
\ba
&& \Dl_1(t)\simeq {L_2\over L}\epsilon_{12}\cos\theta_p\Bigl[(1-\cos\tau_{12})
(\bl_p\!\times\!\bl)\!\times\!\bl\nonumber\\
&&\qquad\quad ~~~ +\sin\tau_{12}\, (\bl_p\!\times\!\bl)\Bigr],
\label{eq:deltal1}\ea
where we have used $\bl\cdot\bl_p=\cos\theta_p$ and
$\tau_{12}\equiv (\omega_{12}+\omega_{21})t$.
Using Eqs.~(\ref{eq:deltal1}) and (\ref{eq:l1l2}), we then find that the mutual inclination
angle $\theta_{12}$ between $\bl_1$ and $\bl_2$ is given by
\be
\left|\sin\theta_{12}\right|=\left|\bl_1\times\bl_2\right|
\simeq
\epsilon_{12}\left|\sin 2\theta_p \sin{\tau_{12}\over 2}\right|.
\label{eq:os1}\ee
Thus, the maximum and the RMS values of $\left|\sin\theta_{12}\right|$ are 
\ba
&&\left|\sin\theta_{12}\right|_{\rm max}\simeq \epsilon_{12}\left|\sin 2\theta_p\right|,
\label{eq:th12max}\\
&&\left\langle\sin^2\!\theta_{12}\right\rangle^{\!1/2}\simeq
{1\over\sqrt{2}}\epsilon_{12}\left|\sin 2\theta_p\right|.
\label{eq:th12max2}
\ea

\subsection{Weak Coupling Limit: $\epsilon_{12}\gg 1$}

In the {\it weak coupling limit}, with $\epsilon_{12}\gg 1$, the vectors
$\bl_1$ and $\bl_2$ precess around $\bl_p$ independently, with constant $\bl_1\cdot\bl_p\simeq
\bl_2\cdot\bl_p\simeq\cos\theta_p$.
Thus
\be
\left|\bl_1\times\bl_2\right|^2
\simeq \sin^2\!2\theta_p\,\sin^2\!\Bigl({\Delta\phi_{12}\over 2}\Bigr)
+\sin^4\!\theta_p \,\sin^2\!(\Delta\phi_{12}),
\label{eq:osc2}\ee
where $\Delta\phi_{12}\simeq (\Omega_{2p}-\Omega_{1p})(\cos\theta_p) t$.
The maximum of $\theta_{12}$ and the RMS value of $\left|\sin\theta_{12}\right|$ are 
\ba
&&(\theta_{12})_{\rm max}\simeq 2\theta_p, \label{eq:the12max}\\
&&\left\langle\sin^2\!\theta_{12}\right\rangle^{\!1/2}\simeq
{1\over\sqrt{2}}\left(\sin^2\!2\theta_p+\sin^4\!\theta_p\right)^{1/2}.
\label{eq:the12max2}
\ea

\subsection{Resonance}

Figure 2 reveals that when $m_2\go m_1$ (i.e., the outer planet is more massive
than the inner planet), a resonance feature appears around
$\epsilon_{12}\sim 1$. At the resonance, $(\theta_{12})_{\rm max}$ can
become much larger than the weak-coupling limit, $2\theta_p$.
When $m_1\go m_2$, no resonance feature exists.

\begin{figure}
\centering
\includegraphics[scale=0.43]{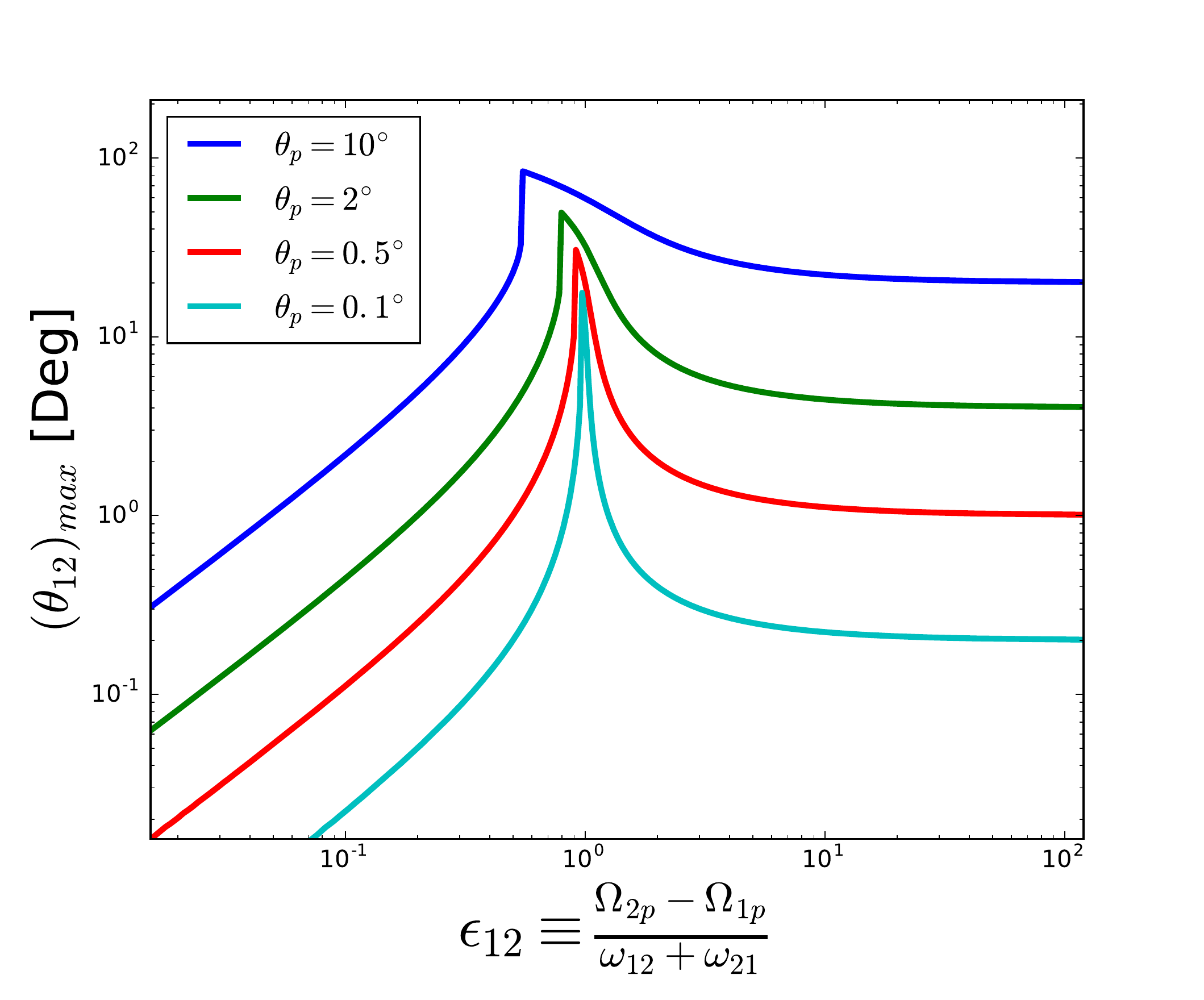}
\caption{Maximum misalignment angle between two inner planets in the presence of 
an external perturber, as a function of $\epsilon_{12}$, in the limit of $m_2\gg m_1$.
Note that in this limit, $\omega_{21}=(L_1/L_2)\omega_{12}\ll \omega_{12}$.
The two planets are 
located at $a_1=0.3$~au and $a_2=0.5$~au. Different curves correspond to different
inclination angles ($\theta_p$) of the external perturber.
These curves are obtained analytically by solving Eq.~(\ref{eq:thjdmax}) derived
in Appendix A. The resonant feature is most prominent for $\theta_p\ll 1$ and is located at 
$\epsilon_{12}=1$ in the $\theta_p\rightarrow 0$ limit. As $\theta_p$ increases, the
resonance feature is broadened and shifted to slightly smaller $\epsilon_{12}$.
}
\label{fig3}
\end{figure}

This resonance feature can be understood analytically in the limit 
when the planetary system contains a ``dominant'' planet (labeled ``d'')
which is much more massive than the other planet (labeled ``j''),
i.e., $m_d\gg m_j$. In Appendix A we develop the Hamiltonian theory
for such systems. We show that for $\theta_p\ll 1$, a sharp 
resonance appears at $|\epsilon_{jd}|=1$, or 
\be
\Omega_{dp}=\Omega_{jp}+\omega_{jd},~~ \Bigl({\rm resonance~for}~{m_d\over m_j}\gg 1,\,
\theta_p\ll 1\Bigr).
\label{eq:res-con}\ee
This resonance condition is easy to interpret physically: The dominant planet experiences
nodal precession at the frequency $\Omega_{dp}$ driven by the perturber, while the 
sub-dominant planet $m_j$ precesses at the rate $(\Omega_{jp}+\omega_{jd})$ driven by
both the perturber and the dominant planet; resonance occurs when these two
precession frequencies match\footnote{
The resonance can also be ``visualized'' geometrically (see Fig.~2 in Lai 2014)
by considering Eq.~(\ref{eq:dlrot}) with $1\rightarrow j$, $2\rightarrow d$,
$\bl\rightarrow \bl_d$ and $\Omega_L\rightarrow\Omega_{dp}$: When $\bl_p$, $\bl_d$ and
$\bl_j$ are approximately aligned, and when the system is near resonance, 
$\bl_1$ precesses around the vector $(\bl_p-\bl_d)$, which is almost perpendicular to 
$\bl_1$, thus producing a large $\theta_{jd}$.}.
Clearly, to satisfy Eq.~(\ref{eq:res-con}) requires
$\Omega_{dp}>\Omega_{jp}$, or $a_d>a_j$, i.e., the dominant planet exterior to the
sub-dominant planet. Near the resonance, the maximum mutual inclination behaves as
(see Appendix A)
\be
(\theta_{jd})_{\rm max}\simeq {2\epsilon_{jd}\theta_p\over |\epsilon_{jd}-1|}
\ee
(valid for general $\epsilon_{jd}$ but $\theta_p,\theta_{jd,{\rm max}}\ll 1$).
This provides an estimate for the ``width'' of the resonance for $\theta_p\ll 1$)..
As $\theta_p$ increases, the resonance becomes broader and is shifted slightly to 
smaller $\epsilon_{12}$ (see Fig.~\ref{fig3}).  Also, as the mass ratio $m_d/m_j$ decreases,
the resonance feature gradually become ``smoothed'' out and disappears
when $m_d/m_j\lo 1$ (see Fig.~\ref{fig2}).

\section{Multi-Planet Systems with External Perturber}

The evolution equations for the orientations of multi-planet ($N>2$) systems
with an external perturber can be easily can be generalized (see Appendix B).
Figures \ref{fig4}-\ref{fig6} show some numerical results for a 4-planet system
($N=4$) in the presence of an external perturber.
To characterize the mutual misalignment of the planets for a wide range of
parameters, we take the dominant planet (the one with the largest mass, labeled ``d'') 
in the system and measure the relative inclination ($\bl_j$) 
of the other planets with respect to $\bl_d$. We define the RMS of $|\bl_j\times\bl_d|$
as 
\be
{\rm RMS}\left(\sin\!\Delta\theta\right)\equiv 
\left({1\over N-1}\Bigl\langle\sum_{j\ne d}\bigl|\bl_j\times\bl_d\bigr|^2\Bigr\rangle
\right)^{\!1/2},\label{eq:rms-define}\ee
and the mutual inclination spread as
\be
\sigma_\theta\equiv \sin^{-1}\Bigl[{\rm RMS}\left(\sin\!\Delta\theta\right)\Bigr].
\label{eq:sig-define}\ee
We also define the averaged coupling parameter of the system as
\be
\bar\epsilon\equiv \left({1\over N-1}\sum_{j\ne d} |\epsilon_{jd}|^2\right)^{1/2}.
\label{eq:bareps}\ee
Other ways of characterizing mutual inclinations are possible (see Appendix B), 
but Eqs.~(\ref{eq:rms-define}) and \ref{eq:bareps}
allow for simple analytical expressions
in the limiting cases, as we discuss below.

\begin{figure}
\hskip -18pt
\includegraphics[scale=0.23]{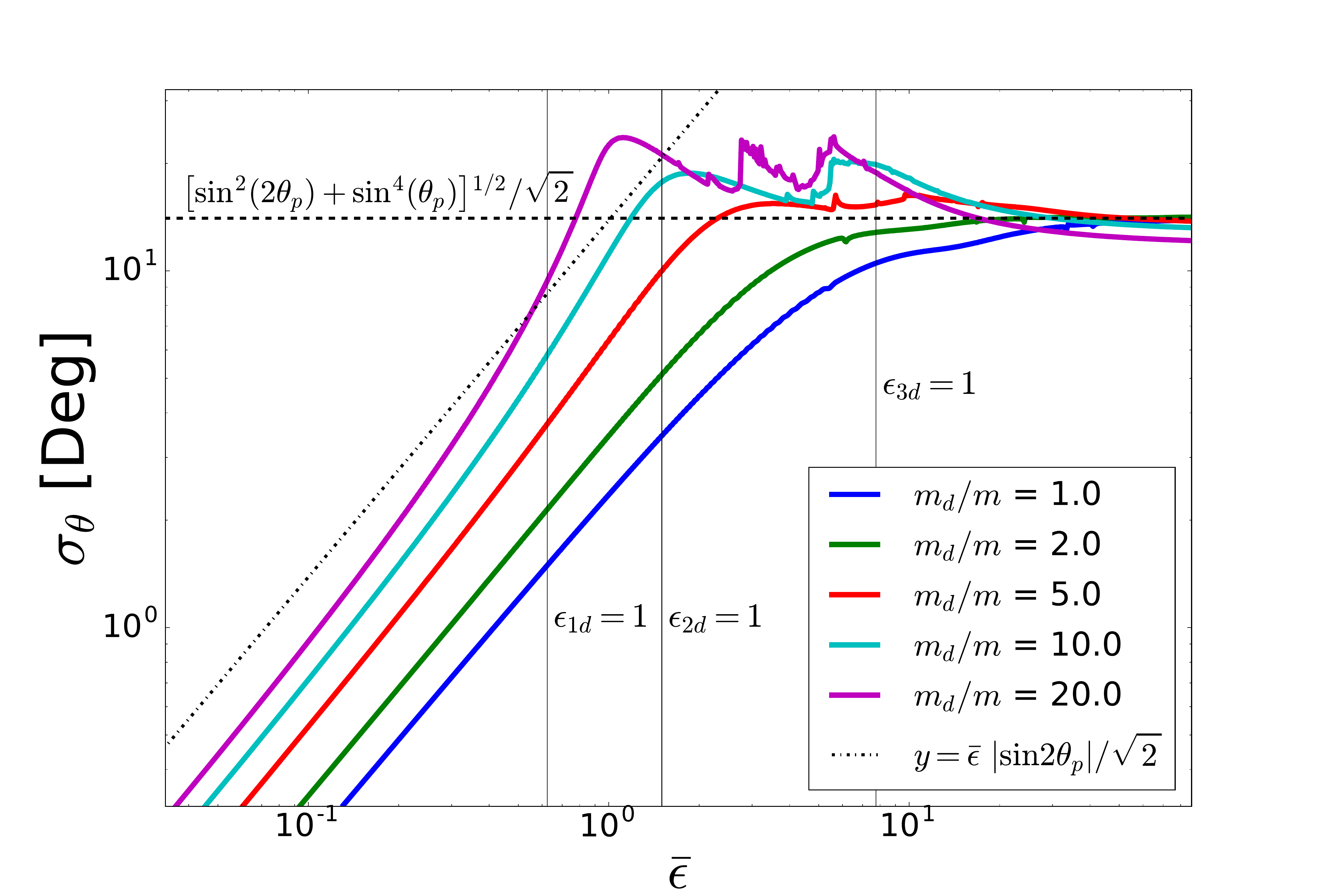}
\caption{Spread in mutual inclination $\sigma_\theta$ (defined
by Eqs.~\ref{eq:rms-define}-\ref{eq:sig-define}) between four planets
in the presence of an external perturber ($m_p$) as a function of the coupling
parameter $\bar\epsilon$ (defined by Eq.~\ref{eq:bareps}).
The four planets are located at the semi-major axes $0.1,~0.15,~0.25,~0.4$~au, and
are initially coplanar and inclined relative to the perturber 
at $\theta_p=10^\circ$. The dominant planet (the one with the largest mass, $m_d$) 
is the 4th planet (with the largest semi-major axis),
and the other three planets have the same mass $m_j=m$. 
The different curves correspond to different mass ratio $m_d/m$, as indicated. 
The dimensionless parameter $\bar\epsilon$ is varied by varying the
``strength'' of the perturber, $m_p/a_p^3$.
Analytical results in the strong coupling limit (Eq.~\ref{eq:strong-coup})
and weak coupling limit (Eq.~\ref{eq:weak-coup}), derived under the assumption $m_d/m\gg 1$,  
are also shown. Three resonance features are present 
when $m_d/m$ is sufficiently large. In the limit of $m_d/m\gg 1$ and
$\theta_p\rightarrow 0$, these resonances are located at $\epsilon_{1d}=1$,
$\epsilon_{2d}=1$ and $\epsilon_{3d}=1$.
}
\label{fig4}
\end{figure}

\begin{figure}
\hskip -18pt
\includegraphics[scale=0.23]{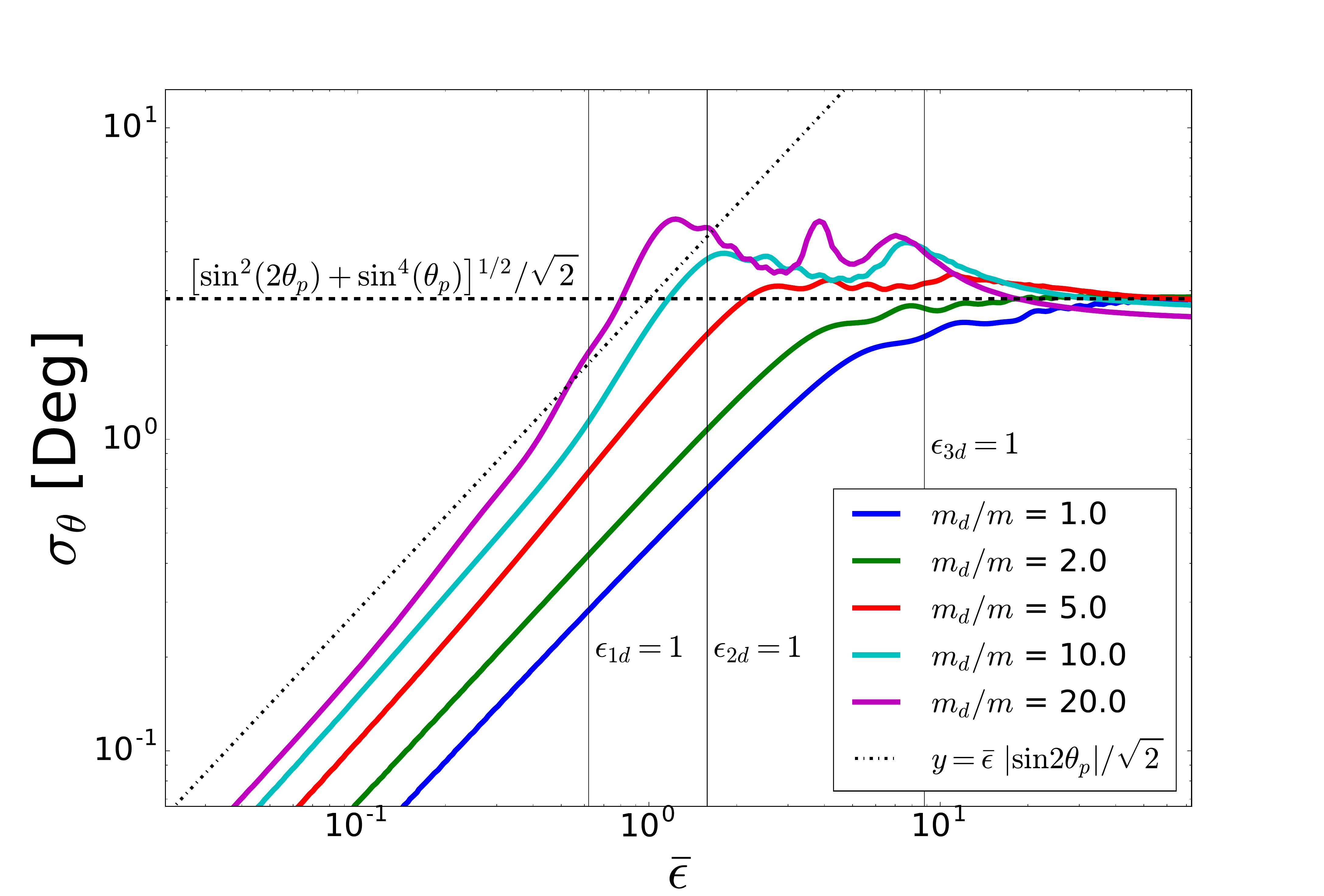}
\caption{Same as Fig.~4, except for $\theta_p=2^\circ$.
}
\label{fig5}
\end{figure}

\begin{figure}
\hskip -18pt
\includegraphics[scale=0.23]{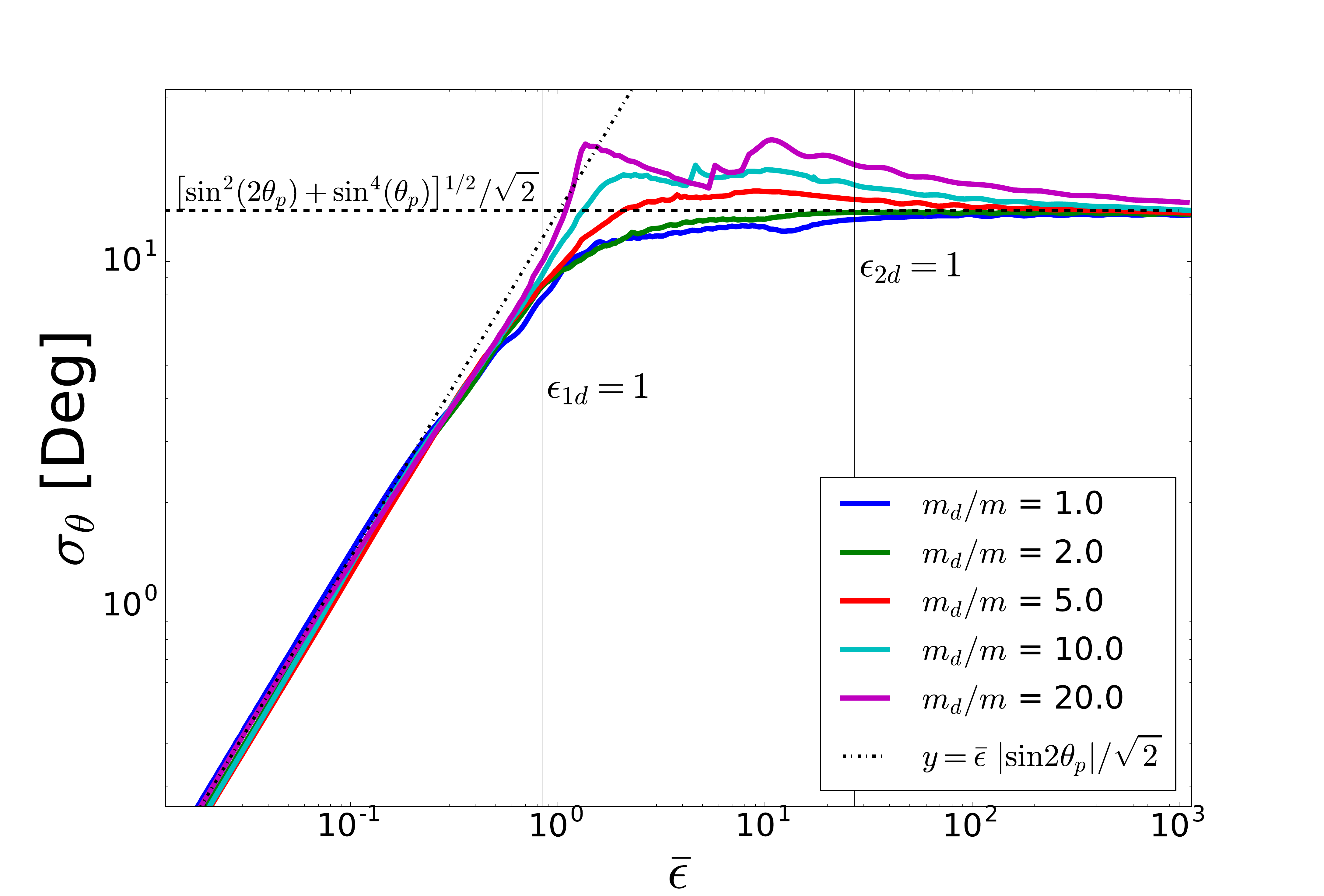}
\caption{Same as Fig.~4, except that the third planet is the dominant planet.
}
\label{fig6}
\end{figure}

To understand the numerical results shown in Fig.~\ref{fig4}-\ref{fig6},
we consider the limiting 
case of $m_d\gg m_j$ (with $j\ne d$).
The angular momentum axis of the
dominant planet precesses around $\bl_p$ with an approximately constant $\bl_d\cdot
\bl_p=\cos\theta_p$. The sub-dominant planets are ``shepherded'' by $m_d$ in 
addition to the external perturber $m_p$. 

In the {\it strong coupling limit}, with $|\epsilon_{jd}|\ll 1$, where
\be 
\epsilon_{jd}\equiv
{\Omega_{dp}-\Omega_{jp}\over\omega_{jd}+\omega_{dj}},
\label{eq:epsjd}\ee
we have (cf.~Eq.~\ref{eq:deltal1})
\ba
&& \bl_j-\bl_d\simeq \epsilon_{jd}\cos\theta_p\Bigl[(1-\cos\tau_{jd})
(\bl_p\!\times\!\bl_d)\!\times\!\bl_d\nonumber\\
&&\qquad\quad +\sin\tau_{jd}\, (\bl_p\!\times\!\bl_d)\Bigr],
\label{eq:deltalj}\ea
where 
$\tau_{jd}\equiv (\omega_{jd}+\omega_{dj})t$. Thus
\be
\left|\sin\theta_{jd}\right|=\left|\bl_j\times\bl_d\right|\simeq 
\left|\epsilon_{jd}\sin 2\theta_p \sin{\tau_{jd}\over 2}\right|.
\ee
The misalignment spread of the $N$ planets is measured by 
\be
{1\over N-1}\Bigl\langle\sum_j\bigl|\bl_j\times\bl_d\bigr|^2\Bigr\rangle
\simeq 
{1\over 2(N-1)}\Bigl(\sum_j |\epsilon_{jd}|^2\Bigr)\sin^2\!2\theta_p,
\ee
i.e., 
\be
{\rm RMS}\left(\sin\!\Delta\theta\right)\simeq
{1\over\sqrt{2}}\,{\bar\epsilon}\, \left|\sin 2\theta_p\right|.
\label{eq:strong-coup}\ee
From Eq.~(\ref{eq:deltalj}), we also find
\be
\langle\sin^2\!\theta_{jk}\rangle=
\Bigl\langle|\bl_j\times\bl_k|^2\Bigr\rangle
\simeq {\sin^2\!2\theta_p\over 2}\left(\epsilon_{jd}^2+\epsilon_{kd}^2-\epsilon_{jd}\epsilon_{kd}
\right).
\ee
In Appendix B we present a more rigorous way to characterize the mutual inclination
and the exact analytical expression in the strong coupling limit.


In the {\it weak coupling limit}, with $|\epsilon_{jd}|\gg 1$, all $\bl_j$'s precess
independently around $\bl_p$. We have
\be
{\rm RMS}\left(\sin\!\Delta\theta\right)
\simeq {1\over\sqrt{2}}\left(\sin^2\!2\theta_p+\sin^4\!\theta_p\right)^{1/2},
\label{eq:weak-coup}\ee
and 
\be
\left\langle\sin^2\!\theta_{jk}\right\rangle^{\!1/2}\simeq
{1\over\sqrt{2}}\left(\sin^2\!2\theta_p+\sin^4\!\theta_p\right)^{1/2}.
\ee

Figures~\ref{fig4}-ref{fig6} show that the numerical results match the
analytical expressions in both strong and weak coupling limits.
\footnote{Note that in the strong-coupling regime,
the agreement between the numerical result and the analytical expression is much better
in Fig.~\ref{fig6} than in Figs.~\ref{fig4} and \ref{fig5}.
The reason is that putting the dominant planet in the middle of the system
makes it more ``dominant'' because of its stronger coupling with
the other inner planets, whereas placing the dominant planet at the
edge of the inner system makes it not as dominant in terms of mutual
couplings (i.e. some of the farther planets may have comparable
coupling to each other compared to the ``dominant'' planet in this case).}
Resonance features also occur whenever a ``minor'' planet
exists inside the dominant planet. The resonance is located at
$\epsilon_{jd}\sim 1$ (with $a_j<a_d$).

Note that the small-scale non-smooth features seen in Figs.~\ref{fig4}-\ref{fig6} are
real, and likely result from the chaotic behavior the system due to 
the overlap of multiple nonlinear resonances. We will study this and other 
related issues in a future paper.

\section{Summary and Discussion}

\subsection{Key Results}

We have calculated the excitation of mutual inclinations in compact planetary systems
by external planetary or stellar companions. Our key results are summarized in 
Figs.~\ref{fig2}-\ref{fig6} and a number of approximate analytic expressions
can be used to assess the importance of external perturbers of various masses ($m_p$), 
semi-major axis ($a_p$) and inclination ($\theta_p$). In general, the mutual
inclination excited by a perturber depends on the dimensionless 
coupling parameter $\epsilon_{12}$ (Eq.~\ref{eq:eps12} or \ref{eq:epsjd}),
which measures the ratio of the differential precession rate of planet $1$ and $2$
induced by the perturber and their mutual precession rate. In order of magnitude,
we have (Eq.~\ref{eq:epsilon12})
\be
\epsilon_{12}\sim 
\left({m_p\over m_2}\right)\left({a_2\over a_p}\right)^3,
\ee
for $m_2\sim m_1$ and $a_2\go a_1$.

For a two-planet system (see Fig.~\ref{fig2}), the mutual inclination
induced by an external companion is comparable to $\theta_p$ when
$\epsilon_{12}\go 1$ (see Eqs.~\ref{eq:the12max}-\ref{eq:the12max2}),
but becomes $\sim \epsilon_{12}\theta_p$ when $\epsilon_{12}\lo 1$
(see Eqs.~\ref{eq:th12max}-\ref{eq:th12max2}). However, when $m_2\go
2m_1$ (i.e., the exterior planet is more massive), a resonance feature
appears at around $\epsilon_{12}\sim 1$ where the mutual inclination may
greatly exceed $\theta_p$ (see Fig.~\ref{fig3}). This enhanced
inclination excitation is the resulr of a secular nodal precession resonance
(Appendix A).

The excitation of mutual inclinations in systems with more planets is
necessarily more complex (Section 3 and Appendix B). Nevertheless,
qualitative similar results can be obtained when the mutual
inclination is measured relative to the more massive (``dominant'')
planet in the system and when an averaged coupling parameter
$\bar\epsilon$ is introduced (Eq.~\ref{eq:bareps}).  Indeed, our
approximate analytic expressions for the mutual inclination spread
(Eq.~\ref{eq:strong-coup} in the strong coupling limit and
Eq.~\ref{eq:weak-coup} in the weak coupling limit) are in agreement
with the numerical results (see Figs.~\ref{fig4}-\ref{fig6}).

\subsection{Applications to Individual Systems}

As noted in Section 1, a number of ``inner planets + companion''
systems have been observed. Here we discuss some of these systems in
light of our theoretical results.

{\bf Kepler-68} ($M_\star=1.08M_\odot,~R_\star=1.24R_\odot$) has two transiting planets
($m_{1,2}=8.3,~4.8M_\oplus$) at $a_{1,2}=0.0617,~0.0906$~au,
and a non-transiting giant planet $m_p\go 0.95M_J$ at
$a_p=1.4$~au ($e_p=0.18$) (Gilliland et al.~2013). The coupling parameter is 
$\epsilon_{12}\simeq 2.3\times 10^{-3}$ 
using the lower limit for $m_p$. The excited mutual inclination spread of the
two inner planet is $\sigma_\theta=\epsilon_{12}\sin
2\theta_p/\sqrt{2}\lo 0.14^\circ$ (regardless of $\theta_p$),
and is smaller than
$R_\star/a_2=3.6^\circ$, consistent with the coplanarity of the two
inner planets.

{\bf Kepler-48} ($M_\star=0.88M_\odot$, $R_\star=0.89R_\odot$) has
three transiting inner planets ($m_{1,2,3}=0.0124,\,0.046,\,0.015
M_J$) at $a_{1,2,3}=0.053,\,0.085,\,0.23$~au,
and a giant planet ($m_p\go 2.1M_J$) at $a_p=1.85$~au
($982$~days) (Marcy et al.~2014). The coupling parameters are
$\epsilon_{12}\simeq 0.0015$ and $\epsilon_{23}\simeq 0.25$ 
using $m_p=2.1 M_J$.  Significant mutual inclination can be excited between
planet 2 and 3 if $\theta_p$ is large. Requiring $\theta_{23}\sim
\epsilon_{23}\theta_p\lo R_\star/a_3=1.03^\circ$ yields $\theta_p\lo
2.3^\circ$. We therefore predict that the non-transiting planet
(Kepler-48e) is closely aligned with the inner transiting planets.  Note
that since $R_\star/a_p=0.13^\circ$, its transit probability is still small.

{\bf Kepler-56} (with a red giant host star $M_\star=1.32M_\odot$,
$R_\star=4.23R_\odot$) has two transiting planets
($m_{1,2}=0.0695,\,0.57M_J$) at $a_{1,2}=0.103,\,0.165$~au (period
$10.5,\,21$~days). The orbits of the two planets are coplanar within $
\sim R_\star/a_2=6.8^\circ$, and are inclined with respect to the
stellar equator by more than $37^\circ$ (Huber et al.~2013).  
{RV observations reveal a third planet with period 1002~days ($a_p=2.16$~au),
$e_p=0.2$, and $m_p\ge 5.6M_J$ (Oter et al.~2016). 
This implies a coupling parameter of $\epsilon_{12}\ge 1.6\times 10^{-3}$.}
Thus the inner two planets are strongly coupled and their coplanarity is
not affected by any (regardless of $\theta_p$) external perturbers
that satisfy the current RV constraint. However, the observed large stellar obliquity
may require a large $\theta_p$.


{\bf WASP-47} ($M_\star=1.04M_\odot$) contains three transiting
planets (Becker et al.~2015; Dai et al.~2015): a hot Jupiter
($1.16M_J$, $a_2=0.051$~au or 4.16~days) with an inner super-Earth
($1.8R_\oplus$ or $12M_\oplus$, 0.79~days) and an outer Neptune-size
planet ($3.6R_\oplus$ or $10.4\pm 8.4M_\oplus$, 9.03~days). These
inner planets are orbited by an external giant planet ($m_p>1.24M_J$)
with $e=0.13$ and $P=572$~days (Neveu-VanMalle et al.~2016).  The
inner planets are well in the strong coupling regime, with
$|\epsilon_{jk}|\ll 1$ ($\epsilon_{12}=2\times 10^{-4}$ and $\epsilon_{23}
=1.7\times 10^{-4}$).

{\bf Kepler-454} ($M_\star=1.03M_\odot$)
has a 10.6~day ($a_1=0.095$~au) transiting planet ($2.37 R_E$, 6.84$M_\oplus$),
a cold Jupiter ($m_p>4.46 M_J$ at 524~days) and a distant companion
($>$12$M_J$ at $>$10 years) (Gettel et al.~2016). The observed system
has $\epsilon= (m_p/m_1)(a_1/a_p)^3\sim 0.1 (m_p/5M_J)$. A neighboring planet 
$m_2\sim m_1$ would give $\epsilon_{12}\sim (a_2/0.2~{\rm au})^3(m_p/5M_J)$, and
would be easily inclined relative to $m_1$ and not observable.
Thus Kepler-454 could be an example of multi-planet systems that haven been
``disrupted'' by giant planet perturbers.

{\bf GJ 832} ($M_\star=0.832M_\odot$) has a super Earth ($m_1>5.4M_\oplus$ at $0.162$~au)
inside a giant planet ($m_p>0.64M_J$ at $3.4$~au), both discovered by RV
(Wittenmyer et al.~2014). With $\epsilon= (m_p/m_1)(a_1/a_p)^3\sim 0.006 (m_p/M_J)
(m_1/6M_\oplus)^{-1}$, any neighboring planet to $m_1$ is strongly coupled to it.

{\bf 55 Cancri} ($M_\star=0.95M_\odot$) has four inner planets
(e,b,c,f) with $m_{e,b}\simeq 0.027,~0.83M_J$, $m_c>0.17M_J$, $m_f >0.16M_J$ 
at $0.0156,~0.115,~0.24,~0.78$~au, and an external giant planet (d) with
$m_d>3.8M_J$ and $a_d=5.74$~au (Dawson \& Fabrycky 2010).  The outer
planet may be inclined with respect to the line of sight by $\sim
53^\circ$ (McArthur et al.~2004), implying $\theta_p\go 37^\circ$
relative to planet e. Since $\epsilon=(m_d/m_f)(a_f/a_d)^3\sim 0.08
(m_d/5M_J)(m_f/0.16M_J)^{-1}$, planet d, even if highly misaligned,
cannot significantly influence the coplanarity of the inner planets.
The planetary system is also orbited by a distant stellar companion
55 Cnc B at $a_B\sim 1065$~au (projected distance). But this will not perturb
the coplanarity of the planets since $(M_B/m_d)(a_d/a_B)^3\ll 1$.

{ 
{\bf Kepler-444} ($M_\star=0.76M_\odot$) has five sub-Earth radius
planets (0.40 - 0.74$R_\oplus$) at semi-major axes
$0.0418,~0.0488,~0.06,~0.0696,~0.0811$~au (Campante et al.~2015)
orbiting the primary star (Kepler-444A).  Astrometric and RV
observations show that a pair of M dwarfs (BC) with total mass
$m_p\equiv m_{BC}=0.54M_\odot$ orbits around Kepler-444A with
semi-major axis $a_p\simeq 37$~au and eccentricity $e_p=0.864$ (Dupuy et
al.~2016). Both the planetary system and the A-BC binary have edge-on
orbits relative to the line of sight. Using $m_5\sim 0.54M_\oplus$
[from the planet mass-radius relation
$m/M_\oplus\simeq (R/R_\oplus)^{2.06}$], we find the coupling parameter
$\epsilon\sim (m_{BC}/m_5)\left(a_5/a_p\sqrt{1-e_p^2}\right)^3
\sim 0.026$, and thus the five planets are strongly coupled and can maintain
their coplanarity (in agreement with the numerical simulation of Dupuy et al.~2016).
}






\subsection{Implications for Kepler Dichotomy}

The common occurrence of giant planets 
{and stellar companions
outside compact planetary systems (see Sections 1 and 4.2) suggests that 
these giant planets or more massive distant stellar perturbers} can excite mutual 
inclinations in the inner planets, thereby account for an 
appreciable fraction 
of the Kepler ``singles''. Our work provides a quantitative criterion (in terms
of the strength of the perturber, $m_p/a_p^3$) for inclination
excitations.  Continued search for external companions of inner
transiting planets would help constrain various scenarios (see Section 1)
for producing the Kepler dichotomy.

{As noted in Section 1, inclined stellar companions may be a natural consequence 
of the binary formation process, while inclined giant planets may be produced
by strong planet-planet scatterings. In the latter case, the inner multi-planet  
system may experience some excitation of mutual inclinations while the outer
giant planets undergo scatterings. (Of course, if the inner planets are 
not well separated from the outer giants, they may be completely disrupted.)
Our numerical calculations (Pu \& Lai 2016, in prep) suggest that in many cases, 
the mutual inclination excitation in the inner system 
during the outer-planet scattering phase is smaller than the subsequent secular phase.}

In this paper we have focused on the excitation of mutual
inclinations, since they most directly influence the transit
probability of multiple planets.  Eccentricities are also excited by
external companions (Pu \& Lai 2016, in prep).  This may explain why Kepler
``singles'' (or a fraction of them) are more eccentric than the Kepler
``multis'', for which the exists tentative observational evidence
(J. Xie et al. 2016).

While Kepler single-transit systems may contain other planets hidden
from transit observations due to mutual inclinations, it is also
possible that they are true ``singles'' because of the dynamical
influences of external giant planets. For example, when appreciable
mutual inclinations and eccentricities are excited, the inner
planetary systems are likely more unstable and will suffer
self-destruction (e.g. Veras \& Armitage 2004; Pu \& Wu 2015). In
addition, as noted above, the inner planetary systems could have been
severely disrupted while strong planet scatterings took place at a few
au's that produced inclined/eccentric giant planets. Continued search
for close neighbors of single-transit planets would shed light on this
issue.

\section*{Acknowledgments}

This work has been supported in part by NSF grant
AST-1211061, NASA grants NNX14AG94G and NNX14AP31G, and
a Simons Fellowship to DL from the Simons Foundation.

\appendix

\section{Hamiltonian Theory for Resonance}

We consider a system with a ``dominant'' planet (labeled ``d'')
whose mass and angular momentum are much larger than the other planets
($m_d\gg m_j$ and $L_d\gg L_j$, with $j\ne d$). The Hamiltonian governing
the dynamics of $\bl_j(t)$ is
\be
H=-{1\over 2}\,\omjd\, L_j\, (\bl_j\cdot\bl_d)^2-{1\over 2}\,\Omjp\, L_j 
\,(\bl_j\cdot\bl_p)^2,
\label{eq:Ham}\ee
where we have neglected a non-essential additive constant.
Since $m_d$ is the dominant planet, its $\bl_d$ simply precesses
around $\bl_p$ with a constant rate, $-\Omega_{dp}\,(\bl_d\cdot\bl_p)\simeq
-\Omdp\cos\!\theta_p$:
\be
{\der\bl_d\over \der t}\simeq -\Omega_{dp}(\bl_d\cdot\bl_p)\,\bl_p\times\bl_d.
\label{eq:dld}
\ee
In the frame corotating with $\bl_d$, the Hamiltonian (\ref{eq:Ham}) 
transforms to
\be
H_{\rm rot}\simeq H+\Omega_{dp}\,(\bl_d\cdot\bl_p)\,\bl_p\cdot (L_j \bl_j).
\ee
It is convenient to use the rescaled Hamiltonian,
\ba
&&{\tilde H}_{\rm rot}={H_{\rm rot}\over L_j}\simeq -{1\over 2}\,\omjd\,\cos^2\!\thjd
-{1\over 2}\,\Omjp\,(\bl_j\cdot\bl_p)^2
\nonumber\\
&&\qquad\qquad\quad +\Omdp\cos\!\theta_p\,(\bl_j\cdot\bl_p),
\ea
where
\be
\bl_j\cdot\bl_p=\sin\theta_p \sin\thjd \cos\phjd+\cos\theta_p \cos\thjd.
\ee
Here $\thjd$ and $\phjd$ are the polar angle and azimuthal angle of $\bl_j$
measured relative to $\bl_d$ (i.e., $\cos\thjd=\bl_j\cdot\bl_d$).
Note that $\phjd$ and $\cos\thjd$ form the conjugate coordinate and momentum
for the Hamiltonian ${\tilde H}_{\rm rot}$.

Suppose $\thjd=0$ at $t=0$. Then the phase-space curve for the evolution of
$\cos\thjd$ and $\phjd$ is determined by
\ba
&& 
-{1\over 2}\,\hOmjp\,(\bl_j\cdot\bl_p)^2+\hOmdp\cos\theta_p\,(\bl_j\cdot\bl_p)
+{1\over 2}\,\sin^2\!\thjd \nonumber\\
&&\qquad\qquad \simeq  
\Bigl(\hOmdp-{1\over 2}\hOmjp\Bigr)\cos^2\!\theta_p,
\ea
where
\be
\hOmjp\equiv {\Omjp\over\omjd},\quad
\hOmdp\equiv {\Omdp\over\omjd}.
\ee
The maximum value $\theta_m\equiv (\theta_{jd})_{\rm max}$ 
is achieved at $\phjd=0$ or $\pi$, and is given by
\ba
&& -{1\over 2}\,\hOmjp\,\cos^2(\theta_m\mp\theta_p)
+\hOmdp \cos\theta_p\,\cos (\theta_m\mp\theta_p)\nonumber\\ 
&&\qquad +{1\over 2}\,\sin^2\!\theta_m
\simeq \Bigl(\hOmdp-{1\over 2}\hOmjp\Bigr)\cos^2\!\theta_p,
\label{eq:thjdmax}\ea
where the upper (lower) sign is for $\phjd=0$ ($\pi$).

\begin{figure}
\centering
\includegraphics[scale=0.3]{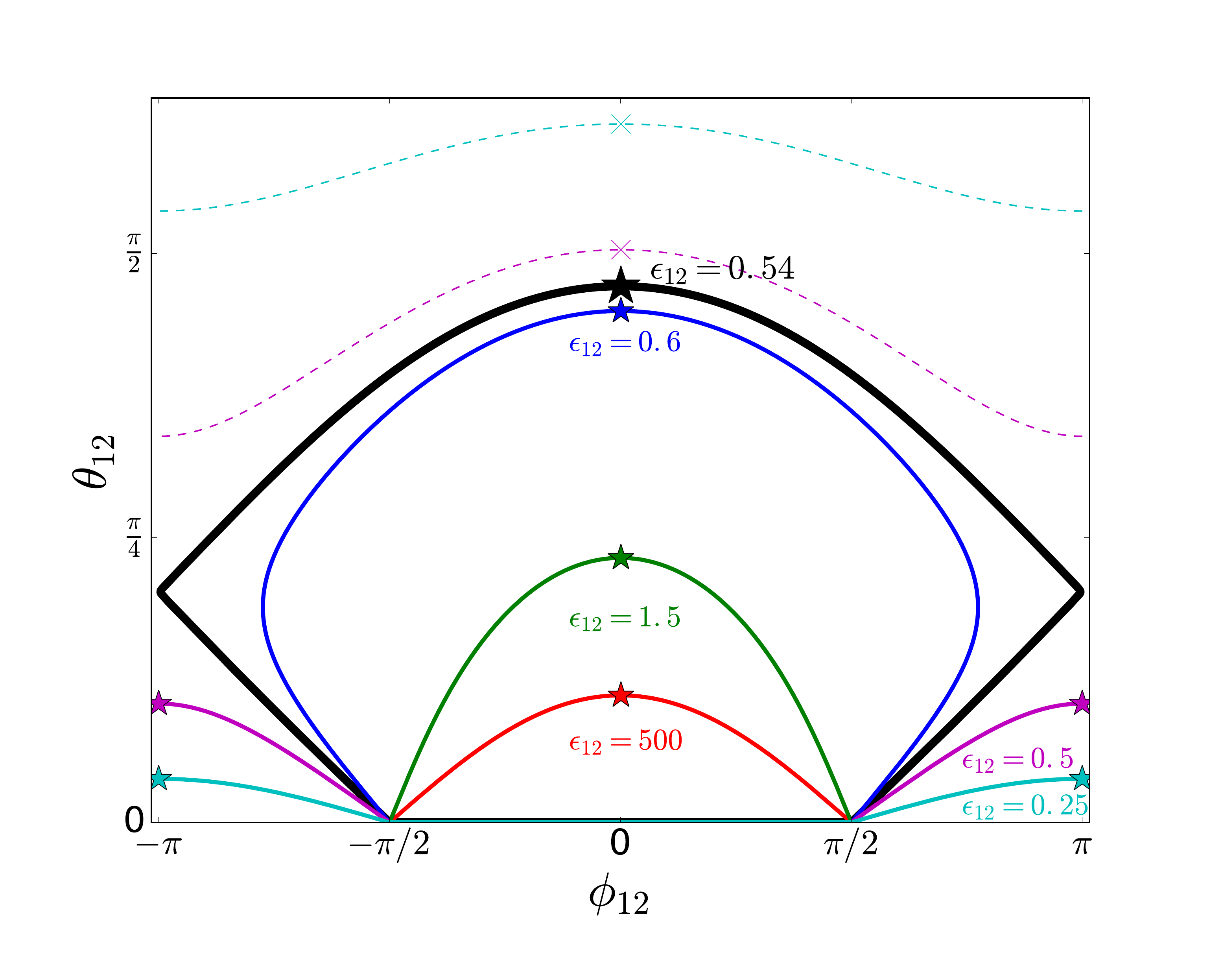}
\caption{
Phase-space curves for the mutual inclination of a two-planet system
with an external perturber. The two planets have $a_1=0.3$~au and $a_2=0.5$~au,
with masses $m_2\gg m_1$, and the perturber's orbit is inclined at $\theta_p=10^\circ$.
The different curves correspond to different values of $\epsilon_{12}$, as indicated;
the solid curves can be reached by an initially aligned system ($\theta_{12}=0$),
while the dashed curves are unreachable. The maximum $\theta_{12}$ for each value of 
$\epsilon_{12}$ is marked. The thick solid line is the separatrix 
(corresponding to a critical value of $\epsilon_{12}$) at which $(\theta_{12}){\rm max}$
experiences a sudden jump (see Fig.~\ref{fig3}).
}
\label{figA1}
\end{figure}

\begin{figure}
\centering
\includegraphics[scale=0.3]{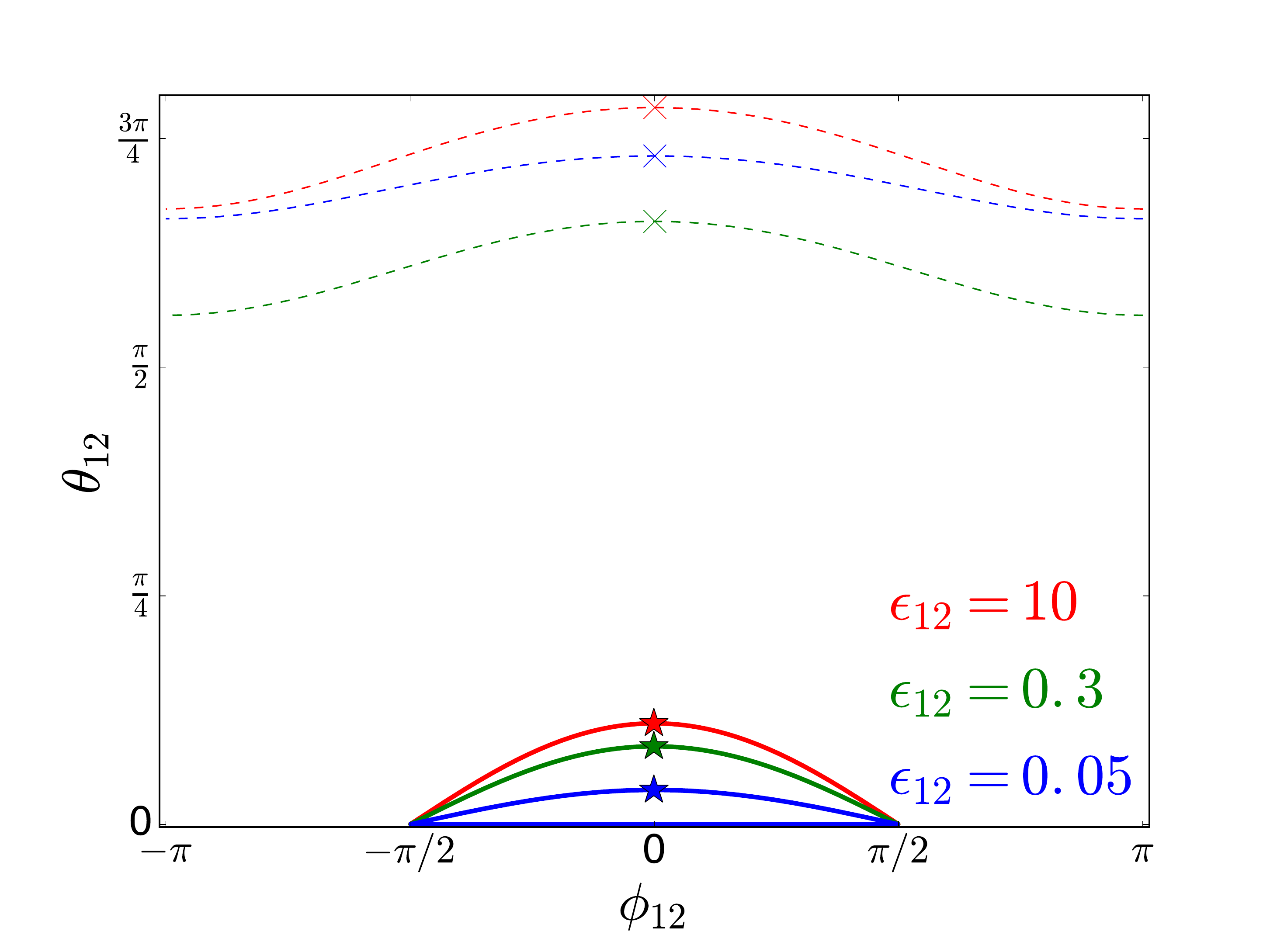}
\caption{
Same as Fig.~\ref{figA1} except for $m_2\ll m_1$.}
\label{figA2}
\end{figure}

Figures~\ref{figA1} and \ref{figA2} show some example phase-space curves for the
cases of $a_d>a_j$ and $a_d<a_j$, respectively. These two cases have very different
phase-space structure, with the former showing clear resonance feature.

Equation (\ref{eq:thjdmax}) can be solved analytically in several limiting cases:

(i) In the {\it strong coupling limit} (but general $\theta_p$), 
$\hOmjp,\hOmdp\ll 1$, we expect $\theta_m \ll 1$. Expanding 
Eq.~(\ref{eq:thjdmax}) for small $\theta_m$, we find
\be
\theta_m\simeq \mp\epsilon_{jd}\sin 2\theta_p,\qquad
{\rm i.e.,}~~\theta_m\simeq |\epsilon_{jd}\sin 2\theta_p|,
\label{eq:theta-strong}\ee
in agreement with Eq.~(\ref{eq:th12max}).

(ii) In the {\it weak coupling limit} (but general $\theta_p$), 
$\hOmjp,\hOmdp\gg 1$, Eq.~(\ref{eq:thjdmax}) has the solution
(see Eq.~\ref{eq:the12max})
\be
\theta_m \simeq 2\theta_p\qquad ({\rm with}~~\phjd=0).
\label{eq:theta-weak}\ee

(iii) In {\it the singular limit of} $\theta_p=0$,  Eq.~(\ref{eq:thjdmax}) has two roots:
The first root is $\theta_m=0$. The second root is 
\be
\cos\theta_m={2\hOmdp\over 1+\hOmjp}-1,\qquad ({\rm 2nd~root})
\label{eq:root2}\ee
which exists only when $|\cos\theta_m| < 1$, or $\epsilon_{jd}=\hOmdp-\hOmjp <1$.

(iv) In the limit of $\theta_p \ll 1$ (but general $\epsilon_{jd}$), the second root
(Eq.~\ref{eq:root2}) remains valid provided that $\theta_m\gg \theta_p$:
\be
\cos\theta_m\simeq {2\hOmdp\over 1+\hOmjp}-1,\qquad ({\rm 2nd~root; valid~for}~\theta_m
\gg \theta_p)
\ee
This root (which exists only when $\epsilon_{jd}=\hOmdp-\hOmjp\lo 1$) cannot be reached
for systems with initially aligned inner planets ($\theta_{jd}=0$) (see Figs.~\ref{figA1}
and \ref{figA2}). The correction to the first root ($\theta_m=0$ in the limit of $\theta_p=0$)
due to finite (but small) $\theta_p$ can be obtained by
expanding Eq.~(\ref{eq:thjdmax}) for $\theta_p,\theta_m\ll 1$. We find
\be
\theta_m\simeq \pm \,{2\epsilon_{jd}\theta_p\over \epsilon_{jd}-1}\qquad
({\rm for~general}~\epsilon_{dj}, ~{\rm but}~\theta_p,\theta_m\ll 1)
\label{eq:theta-res}\ee
(recall that the upper/lower sign is for $\phjd=0,\pi$). Clearly, Eq.~(\ref{eq:theta-res})
reduces to (\ref{eq:theta-strong}) and (\ref{eq:theta-weak}) in the appropriate limits.
Most importantly, Eq.~(\ref{eq:theta-res}) shows that a sharp resonance
occurs when 
\be
\epsilon_{jd}={\Omdp-\Omjp\over \omjd}=1.
\label{eq:res}\ee
At the resonance, $\theta_m\gg\theta_p$ can be attained (but note that 
Eq.~\ref{eq:theta-res} breaks down for $\epsilon_{jd}\rightarrow 1$).
Clearly, the resonance condition can be realized only if $a_d>a_j$ (i.e, the
dominant planet is outside the ``minor'' one). Note that Eq.~(\ref{eq:res})
is exact only in the limit of $\theta_p\rightarrow 0$ and $m_d\gg m_j$;
otherwise the resonance is shifted and broadened (see Figs.~\ref{fig2}-\ref{fig6}).

\section{General Equations for Multi-Planet Systems}

The result of Section 2 can be easily generalized to a system with $N$ inner planets
with an inclined external perturber $m_p$.
The evolution of $\bl_j$ ($j=1,2,\cdots, N$) is governed by the equation
\be
{\der\bl_j\over \der t}=\sum_{k\ne j}
\omega_{jk}(\bl_j\cdot\bl_k)(\bl_j\times\bl_k)
+\Omega_{jp}(\bl_j\cdot\bl_p)(\bl_j\times\bl_p),\label{eq:dlj}
\ee
where
\ba
&&\omega_{jk}={Gm_jm_ka_<\over 4a_>^2 L_j} b_{3/2}^{(1)}\!\left({a_<\over a_>}\right),
\label{eq:ojk}\\
&&\Omega_{jp}={Gm_jm_pa_j\over 4a_p^2L_j} b_{3/2}^{(1)}\!\left({a_j\over a_p}\right),
\label{eq:Ojp}
\ea
with $a_<\equiv \min(a_j,a_k)$ and $a_>\equiv \max(a_j,a_k)$.

In the strong-coupling limit, with $|\bl_j-\bl_k|\ll 1$, the total angular momentum
of the inner binary, $\bL=L\,\bl= \sum_j \bL_j$, evolves according to Eq.~(\ref{eq:dL}),
with the precession rate given by 
\be
\Omega_L\simeq {1\over L}\sum_j L_j\Omega_{jp}.
\ee
In the corotating frame of $\bL$, the evolution of $\Dl_j=\bl_j-\bl$ is governed by the equation
\ba
&& \biggl({\der\Dl_j\over \der t}\biggr)_{\rm rot}\simeq 
-\sum_{k\ne j}\omega_{jk} \bl\times (\Dl_j-\Dl_k)\nonumber\\
&&\qquad\qquad\quad +(\Omega_{jp}-\Omega_L)(\bl\cdot\bl_p)(\bl\times\bl_p).
\label{eq:Deltalj}\ea

We can recast Eq.~(\ref{eq:Deltalj}) into a more convenient form.
Set up a Cartesian coordinate system, with the $z$-axis along $\bl$ and the $y$-axis
along $\bl\times\bl_p$. Let $\Dl_j=(\Delta l_j)_x {\bm {\hat x}} +(\Delta l_j)_y
{\bm {\hat y}}$, and define the complex variable 
\be
I_j\equiv  (\Delta l_j)_x + i (\Delta l_j)_y.
\ee
Then Eq.~(\ref{eq:Deltalj}) reduces to (suppressing the subscript ``rot'')
\be
{\der I_j\over \der t}=- i \sum_{k}A_{jk} I_k + i B_j,
\label{eq:dIj}\ee
where 
\ba
&&A_{jk}=\Bigl(\sum_{n}\omega_{jn}\Bigr)\delta_{jk}-\omega_{jk},\\
&& B_j=(\Omega_{jp}-\Omega_L)\sin\theta_p\cos\theta_p.
\ea
We can write Eq.~(\ref{eq:dIj}) in a more compact form:
\be
{\der\bY\over \der t}=-i\bA\cdot\bY+i\bB,
\label{eq:dY}\ee
where the $N\times N$ matrix has the element $A_{jk}$, and 
\be 
\bY=\left(\begin{array}{c} I_1\\ I_2\\ \vdots\\ I_N\end{array}\right),\quad
\bB=\left(\begin{array}{c} B_1\\ B_2\\ \vdots\\ B_N\end{array}\right).
\ee

In the absence of the external perturber, $B_j=0$, Eq.~(\ref{eq:dIj})
or (\ref{eq:dY}) describes the free inclination oscillations of the $N$-planet system 
(Murray \& Dermott 1999). The eigenmodes $\bY_\alpha$ ($\alpha=1,2,\cdots, N$)
of these free oscillations satisfy the equation
\be
\lambda_\alpha\bY_\alpha=\bA\cdot\bY_\alpha,
\ee
where $\lambda_\alpha$ is the eigenvalue, with $\bY_\alpha\propto \exp(-i\lambda_\alpha t)$.

The general solution of Eq.~(\ref{eq:dY}) takes the form
\be
\bY(t)=\bA^{-1}\!\cdot\!\bB+\sum_\alpha c_\alpha\bY_\alpha \exp(-i\lambda_\alpha t),
\label{eq:yt}\ee
where the constants $c_\alpha$'s are determined by the initial condition.
Assuming $\bY(t=0)=0$, we have
\be
c_\alpha=-\bY_\alpha^\dag\!\cdot\! \bA^{-1}\!\cdot\!\bB,
\ee
where the eigenvector $\bY_\alpha$ has been normalized by 
\be 
\bY_\alpha^\dag\!\cdot\!\bY_\beta=\delta_{\alpha\beta}.
\ee

The mutual inclination in the $N$-planet system is measured by 
\be
{1\over N}\sum_j\bigl|\bl_j\times\bl\bigr|^2={1\over N}\sum_j |I_j(t)|^2={1\over N}
\Bigl |\bY(t)\Bigr|^2.
\ee
Using Eq.~(\ref{eq:yt}), we then have
\be
\left\langle {1\over N}\sum_j\bigl|\bl_j\times\bl\bigr|^2\right\rangle=
{1\over N}\Bigl[\bigl|\bA^{-1}\!\cdot\!\bB\bigr|^2+\sum_\alpha |c_\alpha|^2\Bigr].
\ee


\end{document}